# Local control of a single nitrogen-vacancy center by nanoscale engineered magnetic domain wall motions

Nathan J. McLaughlin[1], Senlei Li[2], Jeffrey A. Brock[3], Shu Zhang[4], Hanyi Lu[1], Mengqi Huang[2], Yuxuan Xiao[3], Jingcheng Zhou[2], Yaroslav Tserkovnyak[5], Eric E. Fullerton[3], Hailong Wang[2], and Chunhui Rita Du[1,2,*]

[1]Department of Physics, University of California, San Diego, La Jolla, California 92093, USA
[2]School of Physics, Georgia Institute of Technology, Atlanta, Georgia 30332, USA
[3]Center for Memory and Recording Research, University of California, San Diego, La Jolla, California 92093-0401, USA
[4]Max Planck Institute for the Physics of Complex Systems, Dresden 01187, Germany
[5]Department of Physics and Astronomy, University of California, Los Angeles, California 90095, USA

*Correspondence to: cdu71@gatech.edu

**Abstract:** Effective control and readout of qubits form the technical foundation of next-generation, transformative quantum information sciences and technologies. The nitrogen-vacancy (NV) center, an intrinsic three-level spin system, is naturally relevant in this context due to its excellent quantum coherence, high fidelity of operations, and remarkable functionality over a broad range of experimental conditions. It is an active contender for the development and implementation of cutting-edge quantum technologies. Here, we report magnetic domain wall motion driven local control and measurements of NV spin properties. By engineering the local magnetic field environment of an NV center via nanoscale reconfigurable domain wall motions, we show that NV photoluminescence, spin level energies, and coherence time can be reliably controlled and correlated to the magneto-transport response of a magnetic device. Our results highlight the electrically tunable dipole interaction between NV centers and nanoscale magnetic structures, providing an attractive platform to realize interactive information transfer between spin qubits and non-volatile magnetic memory in hybrid quantum spintronic systems.

# 1. Introduction

Hybrid quantum structures consisting of state-of-the-art qubits and electronic devices have received immense research interest recently due to their potential for developing next-generation, transformative information technologies.[1–4] Nitrogen-vacancy (NV) centers, optically active spin defects in diamond,[5,6] and magnetic domain walls hosted by solid-state memory devices[7–13] stand out as two promising candidates for use in practical devices. As atomic scale spin qubits, NV centers possess excellent quantum coherence, single-spin addressability as well as unprecedented field and spatial sensitivity.[5,6] Additionally, NV centers provide an attractive platform to develop cutting-edge quantum sensing,[6] network,[14,15] and computing technologies.[2,15–19] On another front, magnetic domain walls in thin films, which sustain nanoscale spatially evolving and reconfigurable spin textures, promise to deliver a wide range of novel functionalities to modern spintronic devices.[11,20–23] Examples include high-speed domain wall-based logic gates,[8,10,24] magnetic racetrack memory,[7,24] long-range, energy-efficient spin transport,[25–27] and many others.[28,29] More recently, magnetic domain walls have been theoretically predicted to enable entanglement between distant spin qubits through dipole-dipole interactions.[12,13]

Despite enormous promise to date, integration of magnetic domain walls with NV centers to realize effective transfer and readout of information encoded in quantum entities and magnetic memory devices remains elusive. One of the major technical challenges involves establishing nanoscale proximity between NV centers and magnetic domain walls in a controllable and reconfigurable way. Here, we show our efforts along this direction. By utilizing NV quantum sensing technologies,[30–35] we achieve nanoscale imaging of spin-orbit-torque (SOT)-induced[9,22,36] domain wall dynamics in Co-Ni multilayer based heterostructures. The internal spin structure of magnetic domain walls is diagnosed by measuring the spatial distribution of the emanating magnetic stray fields. By systematically controlling the SOT-induced domain wall motions, the local field environment of a proximal NV center can be precisely engineered, enabling electrical switching of NV photoluminescence, spin level energies, and coherence time between two different states. Local measurement of NV properties is achieved through the variation of anomalous Hall voltages, which is intimately tied to nanoscale domain-wall-motions in the magnetic channel of the Co-Ni device. Our results demonstrate the two-fold advantages of NV centers in quantum sensing and quantum information science research. The observed electrically tunable coupling between NV centers and propagating magnetic domain walls further highlights the appreciable opportunity for promoting the scalability, quantum interconnection, control of entanglement, and other tailored functionalities of NV-based hybrid quantum systems.[18,37]

# 2. Results and Discussion
## 2.1. Co-Ni magnetic multiplayer device and scanning NV measurement platform

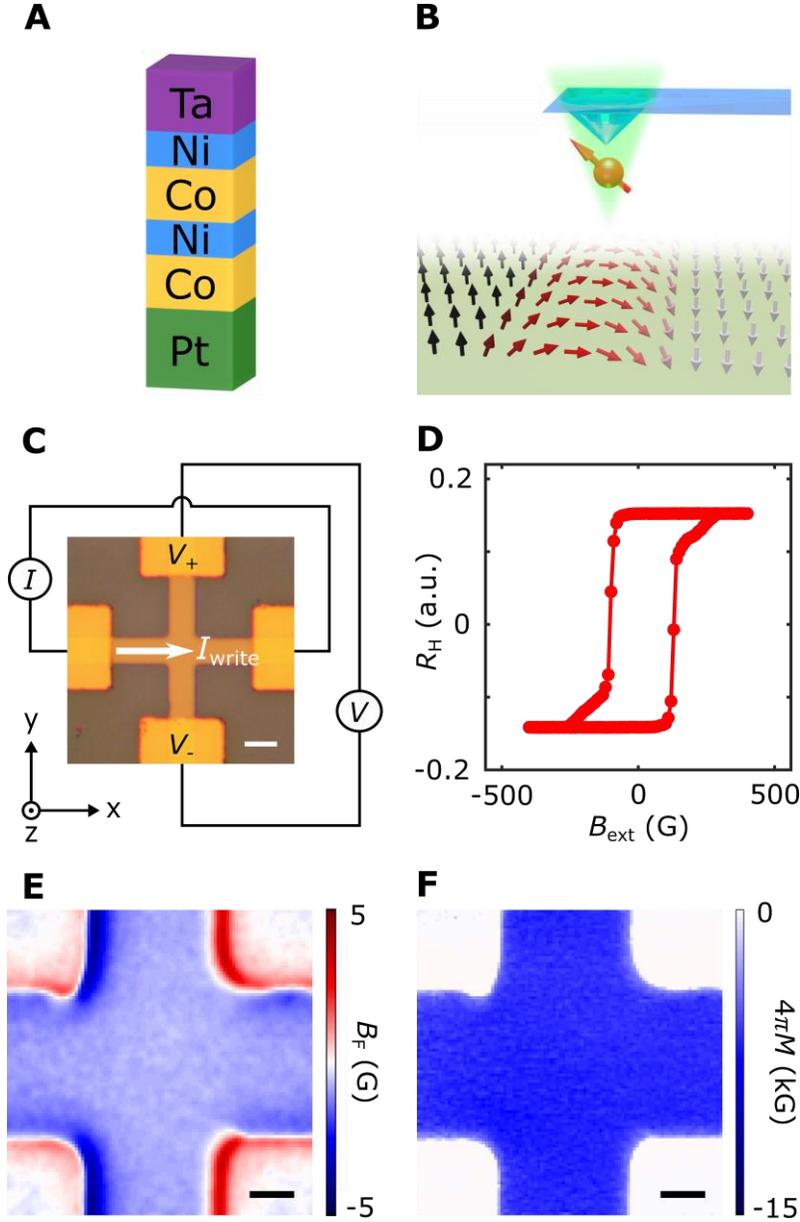

**Figure 1.** Measurement platform and Co-Ni multilayer device layout. (A) Schematic illustration of Co-Ni device structure. (B) Scanning NV magnetometry measurements of a spatially varying spin texture of a Co-Ni multilayer device. (C) Optical microscope image of a patterned Co-Ni multilayer Hall cross device with an illustration of the magneto-transport measurement geometry. Electrical write and read current pulses are applied along the *x*-axis, and the Hall voltage is measured along the *y*-axis. The scale bar is 10 μm. (D) Anomalous Hall resistance $R_H$ of a patterned Co-Ni device measured as a function of an external perpendicular magnetic field $B_{ext}$. (E-F) Two-dimensional (2D) images of magnetic static stray field $B_F$ (E) and reconstructed magnetization $4\pi M$ (F) of a Co-Ni Hall device. The external magnetic field $B_{ext}$ is 30 G applied along the NV axis which is 54 degrees from the out-of-plane direction, and the scale bar is 3 μm.

We first discuss the magnetic films, device structures and our measurement platform as illustrated in Figures 1A-1C. The structure of the magnetic device used in our studies is:

substrate/Pt(5)/[Co(0.5)Ni(0.5)]$_2$/Ta(5) (Section 1, Supporting Information)[38] where the numbers in brackets indicate the thickness of each layer in nanometers. Co-Ni-based multilayer heterostructures were deposited on oxidized Si substrates by magnetron sputtering in a confocal geometry at room temperature. The Co-Ni multilayer has strong perpendicular anisotropy and is sandwiched between a Pt underlayer and a Ta capping layer, both of which have large spin-orbit coupling and a spin Hall effect of opposite sign.[36,38–40] These heavy metal layers serve as efficient and additive spin current sources to drive domain wall dynamics via SOT. The prepared samples were patterned into standard Hall cross structures with a width of ~ 10 μm for electrical transport measurements as shown in Figure 1C. The anomalous Hall characterization (Figure 1D) shows full perpendicular remanence expected for strong perpendicular magnetic anisotropy and thin film thickness.[11,38] To perform nanoscale quantum sensing measurements, a micrometer-sized diamond cantilever[21,30,41] containing an NV single-electron spin is positioned above the surface of the patterned Hall device (Figure 1B). The diamond cantilever is attached to a quartz tuning fork for force-feedback atomic force microscopy measurements. The ultimate spatial resolution of the scanning NV magnetometry system is primarily determined by the NV-to-sample distance,[6,30,33] which stays in the range from 50 nm to 200 nm in our measurements (Section 2, Supporting Information). All the NV measurements presented in the current work were performed at room temperature.

As a first step towards identifying the layout and structure of magnetic domain walls at the nanoscale, we carried out scanning NV imaging of SOT-driven domain wall movements in the Co-Ni multilayer devices. From a microscopic structural viewpoint, an NV center consists of a nitrogen atom adjacent to a carbon atom vacancy in one of the nearest neighboring sites of a diamond crystal lattice. The negatively charged NV state has an $S = 1$ electron spin and acts as a three-level quantum system.[15] Local measurements of stray field $B_F$ arising from the Co-Ni device takes advantage of the Zeeman effect on the NV spin sensor.[5] A static magnetic field along the NV axis will lift the two-fold degeneracy of NV quantum spin states, which can be optically addressed by measuring the spin-dependent NV photoluminescence.[15] The magnitude of $B_F$ can be extracted from the splitting of the NV electron spin resonance (ESR) energy (Section 2, Supporting Information). By scanning the NV center over a mesoscopic length scale above the sample surface, we are able to map the spatially varying $B_F$, enabling nanoscale imaging of local magnetic textures of the Co-Ni multilayer device as shown in Figure 1E. Through established reverse-propagation protocols (Section 3, Supporting Information),[30,33] the corresponding magnetization ($4\pi M$) pattern of the magnetic device can be quantitatively reconstructed, as shown in Figure 1F. The obtained $4\pi M$ of the Co-Ni multilayer sample is ~6763 G, and the strong perpendicular magnetic anisotropy of the sample is evidenced by the uniform spatial distribution of the out-of-plane magnetization.

## 2.2. Quantum imaging of SOT-driven deterministic magnetic switching

We now present data to show nucleation and propagation of domain walls during the SOT-driven deterministic magnetic switching process of the Co-Ni device. In these measurements, millisecond-long electrical write current pulses $I_{\text{write}}$ are applied along the $x$-axis direction of the patterned Hall device, generating spin currents flowing along the ($\pm$)$z$-axis with polarization $s$ oriented along the $y$-axis via the spin Hall effect in the heavy metal Pt and Ta layers,[36,42] as illustrated in Figure 1C. The accumulated spin currents are injected across the Co-Ni/heavy-metal interfaces and exert SOTs on the Néel type domain walls whose chirality is dictated by the interfacial Dzyaloshinskii-Moriya interaction.[38,40,43-46] Application of an in-plane longitudinal magnetic field breaks the energy degeneracy of "up-down" and "down-up" domain walls with respect to the SOT, leading to preferential domain wall motions that accomplish bipolar switching of the Co-Ni magnetization between two magnetic easy states.[38] The anomalous Hall resistance $R_H$ is recorded by supplying a read current with a magnitude of 3 mA after each write current pulse $I_{write}$. Figure 2I shows a typical set of SOT-driven deterministic magnetic switching of the Co-Ni multilayer device. Notably, the measured anomalous Hall signals characterize reversible switching above the positive and negative critical write currents with a polarity depending on the sign of the external bias field (Section 1, Supporting Information), consistent with the mechanism of SOT-

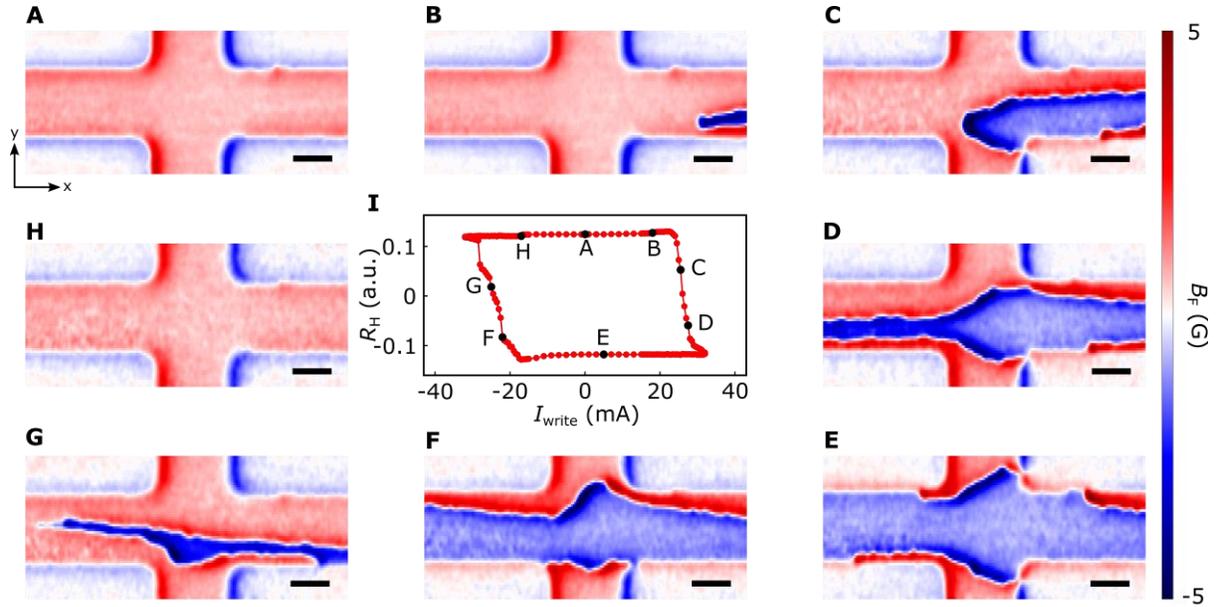

**Figure 2.** Scanning NV imaging of SOT-driven deterministic magnetic switching. (A-H) Nanoscale stray field imaging of domain wall motion during SOT-driven magnetic switching of a Co-Ni device. Scale bar is 5 μm for all images. (I) Anomalous Hall resistance $R_H$ of the Co-Ni multilayer device measured as a function of write current $I_{write}$. For SOT-driven magnetic switching measurements, an external bias magnetic field of 190 G is applied along the applied current direction. Scanning NV imaging results presented in Figures 2A-2H were performed at individual points from "A" to "H" marked on the current-induced magnetic hysteresis loop shown in Figure 2I.

driven magnetic switching.[36]

Next, we utilize NV magnetometry to investigate the formation and propagation of magnetic domain walls at the nanoscale. Scanning NV imaging measurements are performed at the end of each electrical readout current pulse to visualize spin current-induced variations of the local magnetic texture. Figures 2A-2H show a series of representative magnetic stray field $B_F$ maps taken at the corresponding points ("A" to "H") on the SOT-induced magnetic hysteresis loop. At the initial magnetic state "A" where $I_{write} = 0$, the measured magnetic stray field $B_F$ shows a largely uniform distribution over the Hall cross area, indicating a quasi-single domain state of the Co-Ni device with spontaneous perpendicular magnetization. At higher magnitude electrical write current pulses, the effect of the SOT becomes more pronounced, resulting in the nucleation of incipient magnetic domain walls at locations where the energy barrier is lowest, as shown in Figure 2B. The observed magnetic domain wall propagates with increasing write current pulse $I_{write}$ and causes deterministic switching of the Co-Ni magnetization, as shown in Figures 2C-2E. When inverting the polarity of the write current, the up-aligned magnetic domain (along +$z$ direction) is preferentially selected by the exerted SOT, accompanied by the "reversal" of the magnetic switching polarity and the corresponding domain wall motions as shown in Figures 2F and 2G. Lastly, when sweeping the write current to point "H", the Co-Ni multilayer device returns to the initial magnetic state, showing an almost identical stray field map (Figure 2H).

The internal structure of the magnetic domain walls formed in the Co-Ni sample can be diagnosed by measuring the spatial distribution of the emanating magnetic stray fields, which is tied to its underlying spin configuration as illustrated in Figure 3A. Figure 3B shows a zoomed-in view of a stray field map of a formed domain wall in the magnetic device studied. The out-of-plane magnetic moments of Co-Ni multilayer sample exhibit an abrupt spatial rotation, leading to a sign reversal of the measured magnetic stray fields on each side of the domain wall. Figures 3C and 3D plot a line cut of the magnetic stray field $B_x$ and $B_z$ measured by scanning the diamond cantilever across the magnetic domain wall. Our results can be explained by a theoretical model assuming a combination of Bloch and left-handed Néel type magnetic domain walls in the Co-Ni multilayers (Section 4, Supporting Information).[21,47]

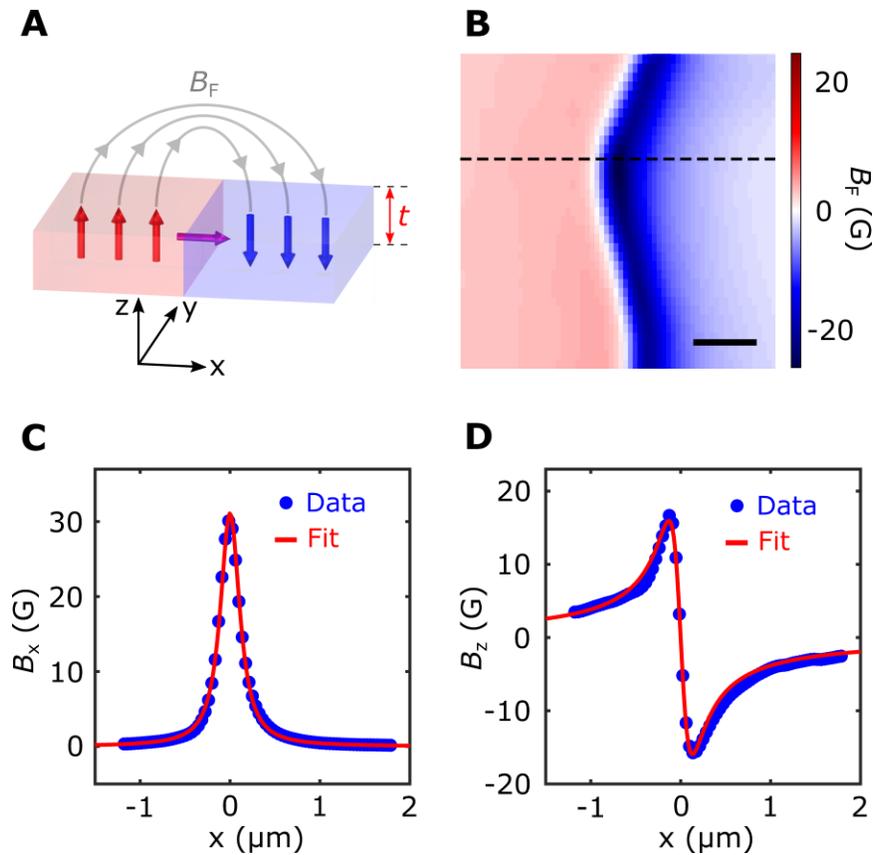

**Figure 3**. Probing the internal spin structure of magnetic domain walls in the Co-Ni multilayer device. (A) Schematic view of a magnetic domain wall formed in a perpendicularly magnetized Co-Ni film. The red (blue) and grey arrows represent the magnetitic moment and emanated stay field, respectively. (B) Scanning NV imaging of stray fields arising from a magnetic domain wall in a Co-Ni multilayer device. Scale bar is 500 nm. (C-D) One-dimensional magnetic stray field $B_x$ and $B_z$ measured along the linecut across the formed magnetic domain wall shown in Figure 3B. The markers are the experimental results, and the solid lines are fittings to a theoretical model assuming a combination of Bloch and left-handed Néel type domain wall.

### 2.3. Local control of a single NV center by domain-wall motions

Our scanning NV magnetometry results highlight highly efficient, current-driven domain wall dynamics at the nanoscale, providing an attractive platform to investigate the dipolar interactions between NV centers and the local magnetic textures of proximal magnetic devices. As such, we demonstrate electrically tunable NV-domain-wall coupling in the presented hybrid system and explicitly show that the photoluminescence, ESR energies, and coherence time of an NV center electron spin can be effectively controlled by reconfigurable domain wall motions. Figures 4A-4B show the schematics and underlying mechanisms of our experiments. The coupling between an NV center and the Co-Ni multilayer sample is mediated by local dipole stray fields arising from the proximal spin textures below the NV sensor. An in-plane, longitudinal magnetic field $B_{ext}$ of 190 Gauss is applied in these experiments, and the distance between the NV center and sample surface is set to be ~59 nm, ensuring sufficiently high NV sensitivity to the variation of local emanating stray fields. When a magnetic domain wall is away from the NV center, the local stray field $B_F$ at the NV site is negligibly small, as illustrated in Figure 4A. In this case, the off-axial component of the external magnetic field $B_{ext}$ will drive the NV center to a mixture of the $m_s = 0$ and $\pm 1$ state with reduced photoluminescence, resulting in a relatively "dark" NV spin state.[48] In contrast, the local field environment is dramatically modified when the domain wall is propagating to the position right underneath the NV center (Figure 4B). A significant in-plane orientated stray field $B_F$, with magnitude of approximately 25-30 G, emerges and effectively compensates the external magnetic field $B_{ext}$, leading to a reduction of the NV off-axis field and enhancement of the NV photoluminescence.

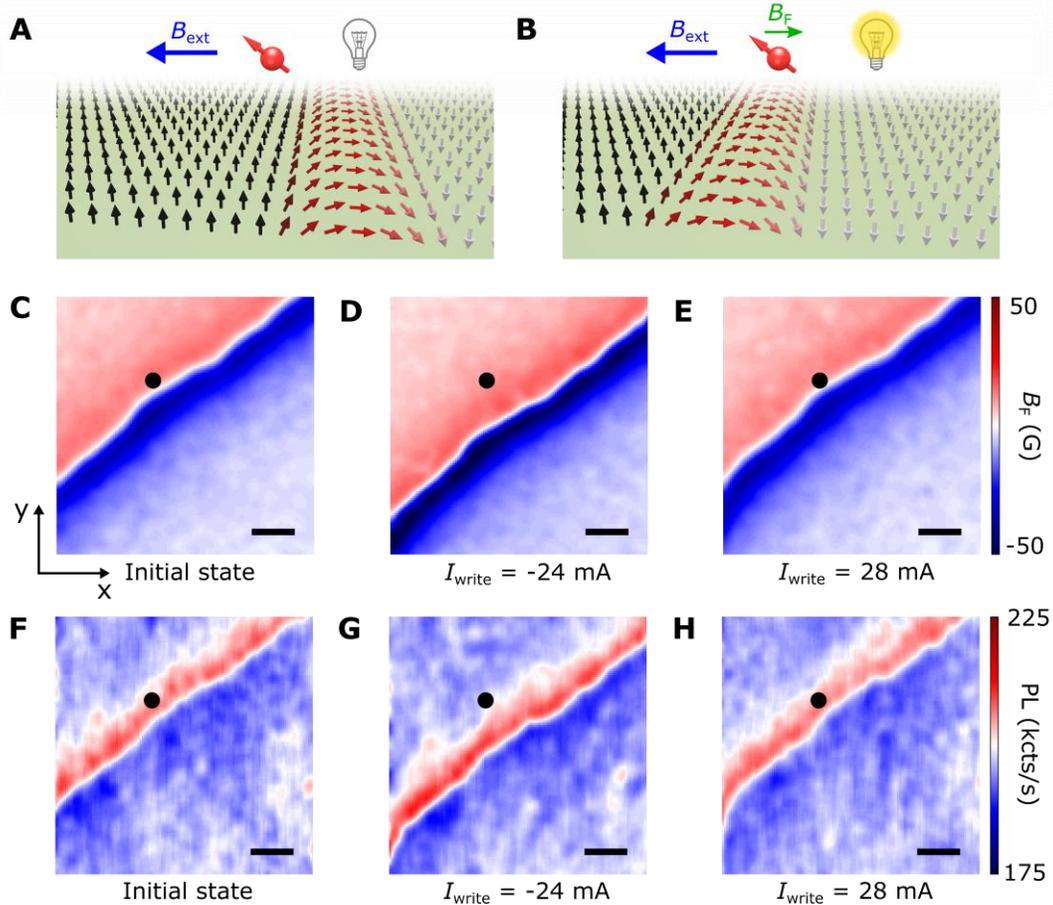

**Figure 4.** Local control of NV photoluminescence by nanoscale engineered magnetic domain wall motion. (A-B) Schematic illustration of tunable dipole interaction between an NV center and a propagating magnetic domain wall, resulting in reconfigurable engineering of the local magnetic field environment at the NV site. (C) Nanoscale stray field imaging of a magnetic domain wall formed in a Co-Ni multilayer device. (D-E) Application of a negative (−24 mA) and a positive (28 mA) electrical write current pulse reversibly drives domain wall forward propagation (D) and backward retraction (E) motions. (F-H) Corresponding NV photoluminescence imaging of the formed magnetic domain wall and its electrically controllable motion. The NV center shows enhanced (reduced) photoluminescence when being positioned right above (away from) the magnetic domain wall. Scale bar is 200 nm for all images. Black points in (C-H) mark the lateral position of the diamond cantilever for local control and measurement of NV properties presented in Figure 5C.

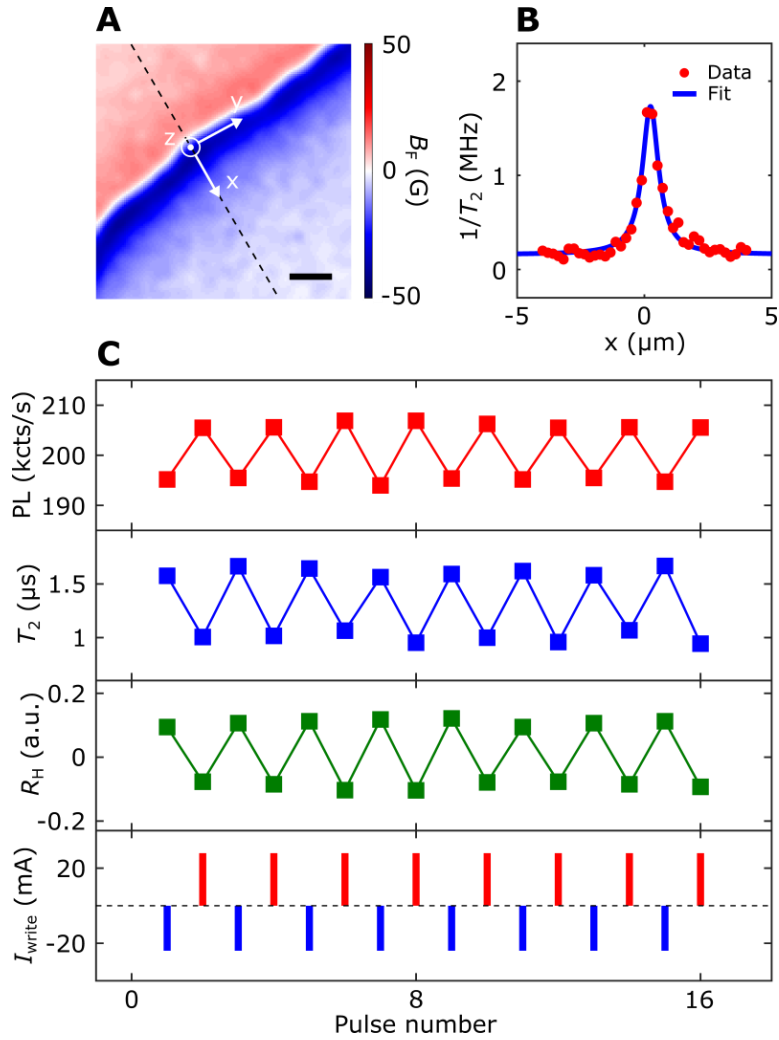

**Figure 5.** Reconfigurable control and measurement of NV photoluminescence and coherence time. (A) Scanning NV imaging of stray field $B_F$ emanating from a magnetic domain wall in a Co-Ni multilayer device. Scale bar is 200 nm. (B) One-dimensional NV spin decoherence rate ($1/T_2$) measured along the linecut (dashed lines) across the formed magnetic domain wall. The experimental results (red points) are in agreement with the theoretical prediction (blue lines). (C) Effective control of NV photoluminescence and coherence time ($T_2$) (top panels) by alternative applications of positive and negative write current pulses (bottom panel). The change of the NV spin properties can be well correlated to the anomalous Hall response driven by reversible domain wall motions in the Hall cross area of the Co-Ni multilayer device. Note that the NV center is not exactly positioned in the center above the formed magnetic domain wall when performing the two-state control measurements.

The scenarios envisioned above are experimentally confirmed by our scanning NV imaging measurements. Figures 4C and 4F present the stray field $B_F$ and photoluminescence maps measured by scanning the NV center over a magnetic domain wall formed in the Co-Ni multilayer device. Notably, the NV center stays in the "bright" state with enhanced photoluminescence when the diamond cantilever is positioned above the magnetic domain wall. As expected, the NV center enters the "dark" state when it is located above the uniform magnetic domain. We note that the lateral positions of the magnetic domain wall can be electrically controlled in a reversible way as shown in Figures 4D and 4E. Application of a write current pulse $I_{write}$ of −24 mA will drive the forward propagation of domain wall over a length scale of ~200 nm. In contrast, sending a current pulse of 28 mA will cause retraction of the magnetic domain wall back to the original position. The measured scanning NV photoluminescence image further confirms the reversible domain wall motions as shown in Figures 4G and 4H. In addition to photoluminescence, ESR energies and the intrinsic quantum coherence time ($T_2$) of the NV spin could also be controlled in a similar way by the nanoscale engineered domain wall motions (Section 5 and 6, Supporting Information). Figure 5B plots the one-dimensional variation of NV spin decoherence rate ($1/T_2$) measured along the linecut (dashed lines) shown in the magnetic stray field map in Figure 5A. The obtained NV $T_2$ shows a significant decrease when the diamond cantilever is scanned across the magnetic domain wall, which is attributed to the gapless magnetic excitation arising from a ferromagnetic domain wall with a divergent susceptibility in the zero-frequency limit.[12,13,18,49] In contrast, the NV center shows an extended $T_2$ when sitting above the uniform magnetic domain. Our experimental results can be well rationalized by a theoretical model (Figure 5B) considering the magnetic noise emanating from a ferromagnetic domain wall (Section 6, Supporting Information).

The observed tunable dipolar interaction between an NV center and a magnetic domain wall provides a new avenue to electrically access the NV properties. When setting an NV center right above a formed magnetic domain wall, alternately applying write current pulses of opposite polarity will reversibly drive the magnetic domain wall either away from or back towards the NV spin, resulting in switching of the measured NV photoluminescence and coherence time between two different states as shown in Figure 5C (Section 5, Supporting Information). By positioning the propagating domain wall in the Hall cross area, the observed binary variation of the NV spin properties can be further correlated to the simultaneously measured anomalous Hall signals of the Co-Ni multilayer device, enabling local control and measurement of NV centers. Note that the measured NV and magneto-transport signals are stable against consecutive write current pulses, demonstrating their robustness against external perturbations. Taking advantage of emergent racetrack memory technologies,[7] we expect that domain-wall-motion driven local addressing of individual quantum states of an array of NV spin qubits could be ultimately achieved.

## 3. Conclusion

In summary, we have demonstrated domain wall motion induced local control and readout of NV photoluminescence, spin level energies, and coherence time in a hybrid quantum spintronic system. By controlling magnetic domain wall motion in a Co-Ni multilayer device, the magnetic field environment of a proximal NV center can be modified in a reconfigurable way, enabling selective writing of NV spin properties. The reproducible control of NV photoluminescence emission and coherence time can be effectively detected by the anomalous Hall response, which is tied to the local magnetic domain wall position in the Hall cross area of the Co-Ni device. Our results illustrate the advantages of NV quantum metrology in studying nanoscale spin behaviors in emergent condensed matter systems. The demonstrated electrically controllable dipole coupling between NV centers and domain walls opens the possibility to realize interactive information transfer between spin qubits and electronic memory devices in hybrid solid-state systems. On a separate note, energy-efficient, current-driven fast domain wall motions may also serve as a suitable, mesoscopic scale quantum transducer/interconnecter,[2,12,37] promoting the functionality of NV centers in developing next-generation, transformative quantum information sciences and technological applications.

**Notes**
The authors declare no competing interests.

**Acknowledgements**. This work was primarily supported by Air Force Office of Scientific Research (AFOSR) under award No. FA9550-20-1-0319 and its Young Investigator Program under award No. FA9550-21-1-0125. Development of scanning NV magnetometry techniques was supported by the U.S. Department of Energy (DOE), Office of Science, Basic Energy Sciences (BES), under award No. DE-SC0022946. Device fabrication and characterization were partially supported by the U. S. National Science Foundation (NSF) under award ECCS-2029558. J. A. B., Y. X. and E. E. F. were supported by the U. S. National Science Foundation (NSF) under award the National Science Foundation, Division of Materials Research Award 2105400. This work was performed in part at the San Diego Nanotechnology Infrastructure (SDNI) of the University of California – San Diego, a member of the National Nanotechnology Coordinated Infrastructure, which is supported by the National Science Foundation Grant ECCS-1542148. The work at UCLA was supported by the US Department of Energy, Office of Basic Energy Sciences under Grant No. DE-SC0012190.

**Data Availability Statement**
The data that support the findings of this study are available from the corresponding authors upon reasonable request.